\documentclass{iau} 
\usepackage{graphicx}

\title[WS 8: Supernovae] 
{Workshop 8: Supernovae}

\author[Maximilian Stritzinger \& Takashi J. Moriya]   
{Maximilian Stritzinger$^1$
 \and Takashi J. Moriya$^2$\thanks{National Astronomical Observatory of Japan (NAOJ) Fellow.}
 }

\affiliation{$^1$Department of Physics and Astronomy, Aarhus University, \\
Ny Munkegade 120, DK-8000 Aarhus C, Denmark \\ email: {\tt max@phys.au.dk} \\[\affilskip]
$^2$Division of Theoretical Astronomy,
              National Astronomical Observatory of Japan, \\
              National Institutes of Natural Sciences,
              2-21-1 Osawa, Mitaka, Tokyo 181-8588, Japan \\email: {\tt takashi.moriya@nao.ac.jp}}

\pubyear{2018}
\jname{Southern Horizons in Time-Domain Astronomy}
\begin{document}

\maketitle

\begin{abstract}
This contribution presents the results of workshop 8 (supernovae) was held  during the IAU Symposium 339 ``Southern Horizons in Time-Domain Astronomy" at Stellenbosch University, South Africa. 
 Workshop 8 covered  a cornucopia of topics with each one having a short presentation by a pre-determined participant followed by a round table discussion. During the first of two sessions, G. Hosseinzadeh and H. Kuncarayakti presented results of their recent works on   interacting supernovae. This includes both the intriguing  Type~Ibn supernova subclass, as well as SN~2017dio, which appears to be the first Type~Ic supernova to exhibit signatures of hydrogen-rich circumstellar interaction at all phases. During the second session, M. Sullivan provided an excellent summary related to the future of transient science in the era of Big Data, and the participants discussed strategies to determine which targets and fields should be selected for spectroscopic followup. Workshop 8 concluded with a rather heated discussion regarding the need for the IAU supernova working group  to consider to modify the current criteria for  a confirmed supernova  to receive an official IAU designation.

\end{abstract}

\firstsection 
\section{Introduction}

The  International Astronomical Union (IAU) Symposium 339 ``Southern Horizons in Time-Domain Astronomy" was  held during week 46, 2017, at the Wallenberg Centre of the Stellenbosch Institute for Advanced Studies (STIAS) in Stellenbosch, South Africa. The meeting provided a timely opportunity for researchers from around the globe to assess the current status of time-domain science. Here we summarize  the science discussions that transpired during  Workshop 8, entitled ``Supernovae". Workshop 8   was fortuitously organized by us over the fortnight leading to the start of the meeting, and discussion topics were chosen to highlight new scientific results with the intent to  inspire discussion on a number of aspects related to the transient Universe. 
We sought to devise an ambitious afternoon supernova session with the format  consisting of a handful of short presentations serving as launching points for further round table discussions. 
Workshop~8  consisted of 2$\times$90 minute sessions with the first session focussing on stripped-envelope supernovae that exhibit signatures of circumstellar interaction, while the second session focused on future surveys and IAU transient classification criteria. 

After welcoming the participants and providing  a quick briefing of the workshop format by  the organisers,   G. Hosseinzadeh kicked off the first session with a  review of the observational properties of today's sample of  Type~Ibn supernovae. This was followed by  H. Kuncarayakti's presentation on the recently studied   Type~Ic supernova (SN) 2017dio,   which  exhibits clear  signatures  of hydrogen-rich circumstellar interaction at all phases of its evolution.
During the second session we shifted gears and discussed the status of transient surveys.   M. Sullivan initiated the session with  an excellent review of past, current and future transient search programs. S. Mattila then presented several examples of highly extinguished transients that are clearly supernovae yet the current IAU criteria prevents them to receive a bonafide IAU supernova designation. This led to a rather heated debate among the workshop participants, and there appeared to be a general consensus  that the  IAU supernova working group should address this untenable situation in the immediate future.

\section{New types of interacting supernovae}

\subsection{Type Ibn supernovae (presented by Griffin Hosseinzadeh)}
Type Ibn supernovae are supernovae interacting with helium-rich circumstellar media. Their spectra are dominated by narrow (FWHM $\sim 2000~\mathrm{km~s^{-1}}$) helium emission lines with little or no hydrogen. After their first identification (SN~1999cq, \cite{matheson00}), many studies on individual objects, including SN~2006jc with a pre-burst (\cite{pastorello07}), have been performed. \cite{hosseinzadeh17} recently reported the observations of six new objects, and these results were presented.  To the best of our knowledge, as of today, the literature sample of Type Ibn supernovae  consists of  $\approx$ 22 objects. 

Light curves of Type~Ibn supernovae are much more homogeneous than those of Type~IIn supernovae that interact with hydrogen-rich circumstellar media. They rise to peak in less than 10~days and decline at a rate of $\sim 0.1~\mathrm{mag~day^{-1}}$. Their decline is also much faster than that of Type~IIn supernovae. Based on their fast evolving light-curve, \cite{moriya16} suggested that the circumstellar media in Type~Ibn supernovae are likely confined near the progenitors. However, curiously the narrow spectral features driven by the interaction appear for weeks or even months post explosion. 

Early spectra of Type~Ibn supernovae can be broadly classified into two types. One type typically exhibits narrow P-Cygni helium profiles superposed on a blue continuum, while the other type show narrow helium emission without P-Cygni absorptions on top of broader features. The spectra of both types evolve to appear similar within $\sim 1~\mathrm{month}$ post explosion. The narrow P-Cygni profiles may result from optically thick circumstellar media, while Type~Ibn supernovae with or without weak P-Cygni profiles, may only have optically thin circumstellar media. However, there can exist observational biases, i.e., all Type~Ibn supernovae may have narrow P-Cygni lines if observed early enough. If the circumstellar material is aspherical, viewing angles could also play an important role in shaping their observational  characteristics. 

All Type~Ibn supernovae appear in star-forming galaxies, except for PS1-12sk (\cite{sanders13}). \textit{Hubble Space Telescope} is slated  to observe the location of PS1-12sk in the near future, in order to search for  star formation activities around the location of this supernova.  

Light-curve properties of Type~Ibn supernovae are very similar to those of rapidly-evolving luminous transients reported by \cite{drout14}. Indeed, some Type~Ibn supernovae show blue continua within a week from peak brightness, while others in the sample might have experienced interaction with helium-rich circumstellar shells without showing clear helium signatures. The fraction of Type~Ibn supernovae is estimated to be only a  few percent of the total core-collapse supernovae rate. If Drout's objects also have helium-rich circumstellar media, almost 10\% of core-collapse supernovae could experience interaction with a dense, helium-rich circumstellar media.

\subsection{A Type Ic supernova interacting with a hydrogen-rich circumstellar medium (presented by Hanin Kuncarayakti)}

This segment of the first session provided the participants with observations of an object that appears to be unique among supernovae.   
SN~2017dio was discovered by the ATLAS survey on 16 April 2017 (UTC) in a faint galaxy at a redshift of $z = 0.037$. 
Upon ignoring narrow emission lines in the spectrum, it was classified as a Type~Ic supernova (\cite{cartier17}). However,  narrow hydrogen and helium emission lines are clearly present in its spectra, which has --up to now-- never been seen in a Type~Ic supernova at early phases. 

\begin{figure}[t]
\begin{center}
 \includegraphics[width=3.4in]{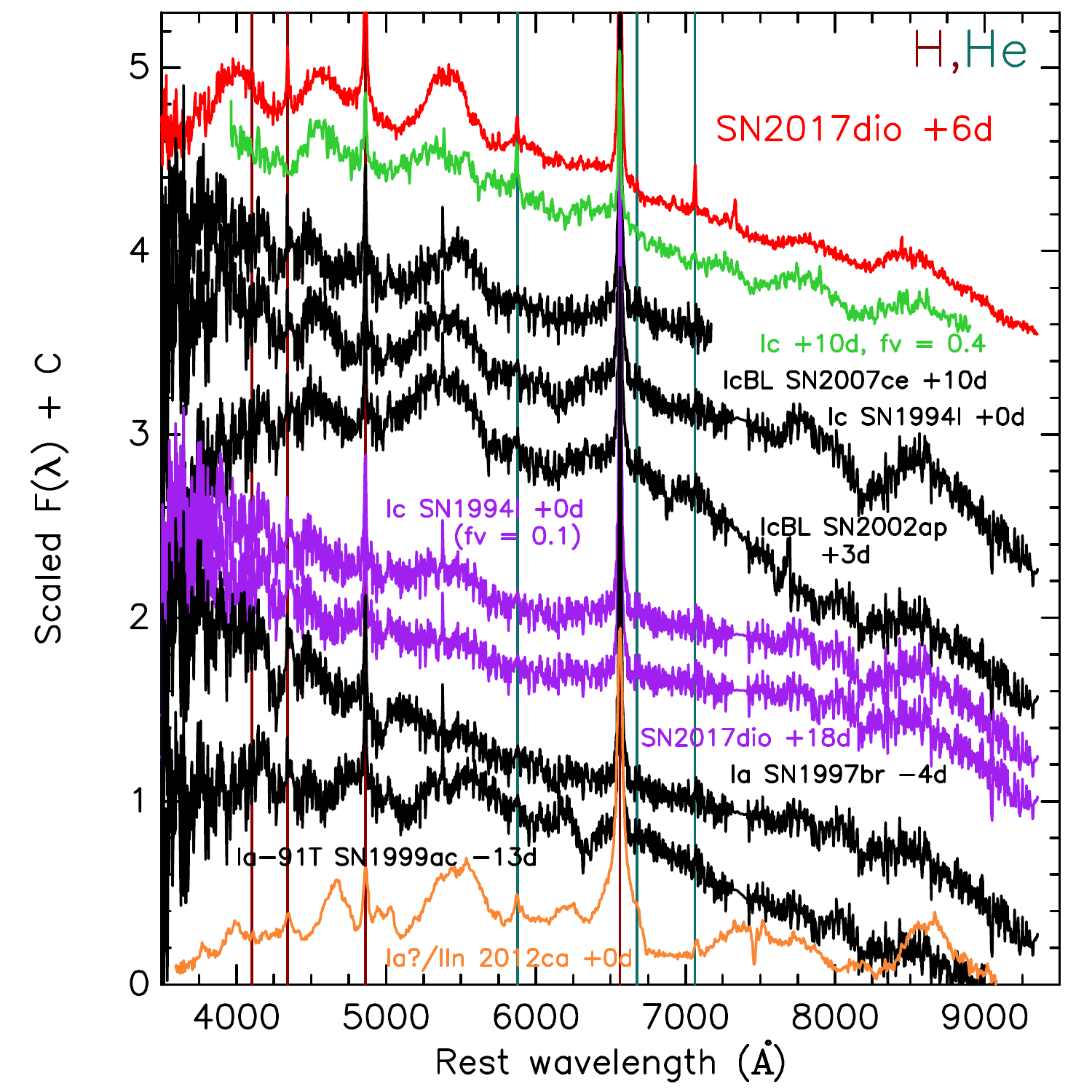} 
 \caption{A spectrum of SN 2017dio compared to those of several types of supernovae coadded with CSM interaction component  (cf. \cite{kuncarayakti18}).}
   \label{fig:sn2017diospec}
\end{center}
\end{figure}

Three spectra observed within 6~days after the discovery are found to be very similar. Then, the spectrum takes on a rather blue and featureless continua, similar to that 
of Type~IIn supernovae. The late-phase spectra  are found to also resemble the spectra of Type~IIn supernovae. Narrow (FWHM $\sim 500~\mathrm{km~s^{-1}}$) hydrogen and helium emission lines are always present in the spectra of SN~2017dio,  suggesting the existence of a dense circumstellar medium. The early spectra are similar to those of Type~Ia supernovae with circumstellar interaction (e.g., SN~2002ic), but the underlying spectra (without the narrow line)  match those of broad-line Type~Ic supernovae  (see Fig.~\ref{fig:sn2017diospec}). 

The light curve of SN~2017dio has distinct features with two peaks. The light curve towards the first peak is well fitted by a Type~Ic supernova template of \cite{taddia15} with an absolute peak $g$-band magnitude of $-$17.5. Then, the luminosity starts to increase again for more than 50~days and reaches a second peak with an absolute $g$-band magnitude of $-$18.8. 
The peak magnitude of the second maximum suggests that there is no Type~Ia supernova hidden below because at optical wavelengths they reach maximum brightest during their first peak, and the first peak of SN 2017dio is too faint for a Type~Ia supernova. Close inspection of the spectra suggests SN 2017dio is likely a Type~Ic supernova, with at least $30-40$\% of its luminosity contribution produced from circumstellar interaction (e.g., \cite{leloudas15}).
\cite{kuncarayakti18} argue that the origin of the circumstellar medium surrounding SN 2017dio is related to significant  mass loss ($\sim 10^{-2}~\mathrm{M_\odot~yr^{-1}}$) from the progenitor system which occurred in the decades preceding its final demise. The asymmetric spectral lines suggest  the circumstellar medium is asymmetric. The big question is then: how does a hydrogen and helium free supernova progenitor obtained a hydrogen-rich circumstellar medium? To find a possible solution we turn to binary evolution to answer this question. A possible binary scenario consists of a primary star that evolves to be a carbon and oxygen star after Roche-lobe overflow. Meanwhile, its secondary companion evolves to a luminous blue variable, or a giant,  driving a second phase of Roche-lobe overflow. The primary star explodes at this moment with a hydrogen-rich circumstellar medium created by the secondary star.

There are  objects that are probably related to SN~2017dio. Type~Ibn supernovae  previous mentioned are obvious. In addition, SN~2014C (\cite{milisavljevic15}) and SN~2001em (\cite{chugai06}) are Type~Ib and Type~Ic supernovae, respectively, that exhibited narrow hydrogen features approximately one year after their explosions, and the existence of detached hydrogen-rich circumstellar media around hydrogen-free supernova progenitors has been suggested. SN~2010mb is another Type~Ic supernova with circumstellar interaction, but it did not show any hydrogen and helium features (\cite{ben-ami14}). Some  superluminous Type~Ic supernovae are known to have late-phase hydrogen emission that begins to appear after a year from the light-curve peak (\cite{yan17}). A few other Type~Ib/Ic supernovae have been suggested to exhibit late-phase hydrogen emission lines (\cite{vinko17}), but SN~2017dio is the \textit{first} Type~Ic supernova that shows   signatures of hydrogen-rich circumstellar interaction right from the beginning.

An interesting question to consider is how many Type~IIn supernovae without early spectra are like SN~2017dio? The late-phase spectra of SN~2017dio are similar to Type~IIn supernovae and it may have been classified as Type~IIn supernovae if the early spectra were not obtained. Other 2017dio-like objects might have previously been misclassified as Type~IIn supernovae. 

\section{Surveys and IAU SN designation}

\subsection{Segment on the history, present and future of transient surveys (presented by Mark Sullivan)}

\begin{figure}[t]
\begin{center}
  \includegraphics[angle=270,width=6.in]{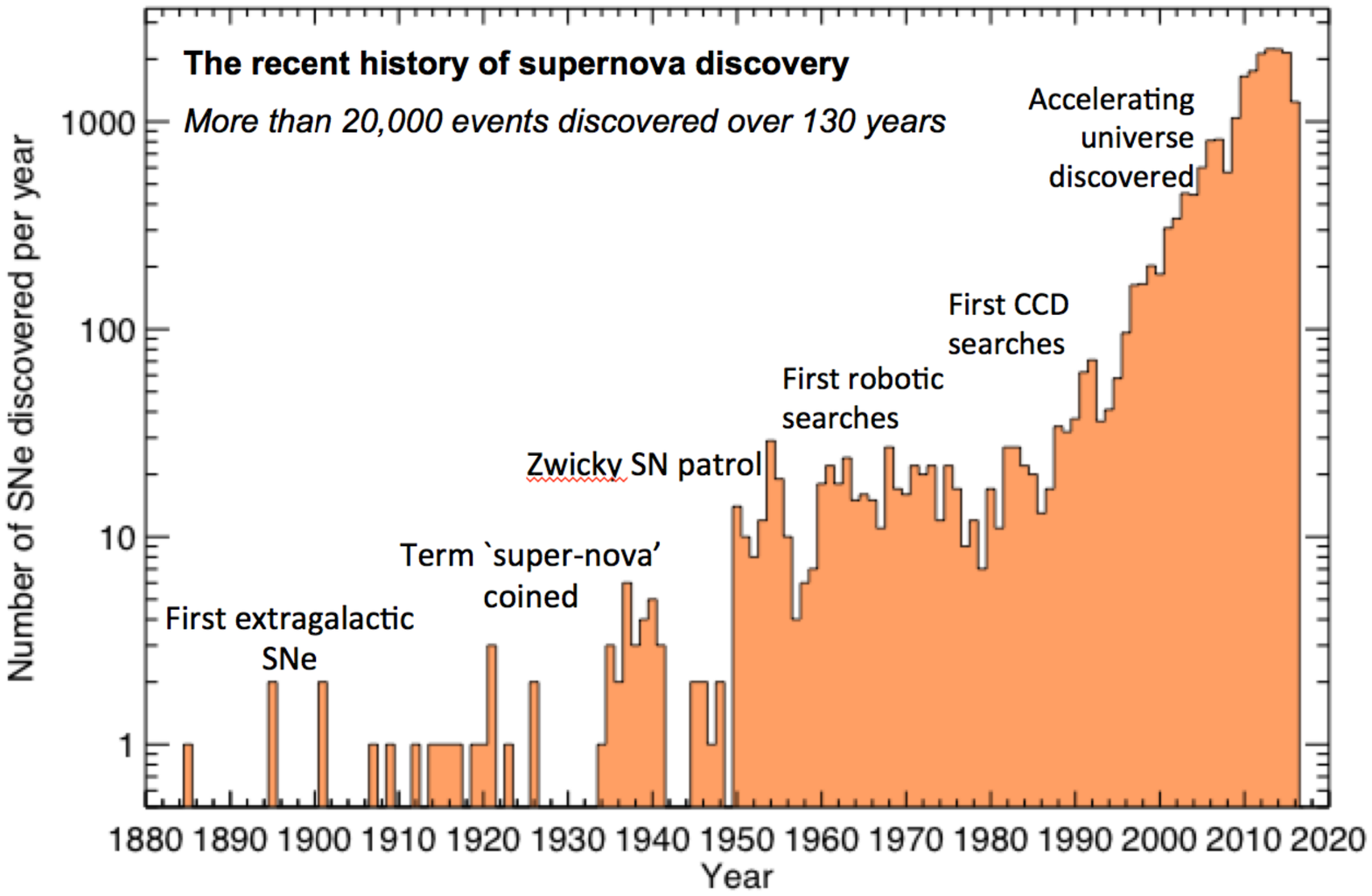} 
\vspace*{-1.5cm}
 \caption{Histogram of supernovae discoveries since the 1880s. Figure courtesy of M. Sullivan. }
   \label{fig:histogram}
\end{center}
\end{figure}

Over the past 130 years (see Fig.~\ref{fig:histogram}), time has witnessed the discovery of the first extragalactic supernova at the end of the 19th century, the coining of the term `super nova' in the 1930s (\cite{baade34}), the emergence of the first robotic searches in the 1960s, followed by the development, implementation and deployment of CCD cameras in the 1980s. 
By late 1990s, with the combination of experience and the advent of  larger and more efficient CCDs,  observations of Type~Ia supernovae lead to the startling discovery that the expansion rate of the Universe is accelerating.  
During the modern era, 20,000 supernovae have been discovered, and between  the advent of the Cal{\'a}n/Tololo survey and extending to today with the Dark Energy Survey,  on average, $\approx$ 5 supernovae are discovered and confirmed per day.

Combined efforts of countless Astronomers has revealed the existence of a multitude of transient objects associated with the terminal demise of their progenitor star(s). Today, researchers in the field study a range of supernova types extending from the very bright to the very faint, however, our knowledge of the transient Universe on short time scales remains largely as unexplored territory. The hope is this will change in the  future with on-going and new high-cadence searches, e.g., ATALS, ASAS-SN, BlackGem, ZTF and  LSST. 

Transient science is in the midsts of sea change with the near future marking the  era of Big Data. For example, LSST is expected to produce 100,000 transient alerts per night based on 20 Terabytes of data. Integrating over its  planned 5 year lifetime, LSST is expected to produce  some 10 million supernovae discoveries! How to manage the study of these objects stands as a significant challenge and solid selection criteria must be developed in order to  determine what objects to study, and this problem will be exacerbated by a lack of spectroscopic followup facilities. Although during his presentation,  M. Sullivan convinced us that constructing a sample  of $\sim$ 30,000 supernovae with spectroscopic confirmation is well within grasp with the realization of  the European Southern Observatory's  VISTA  telescope 4MOST survey, which would build off of experience obtained by the AAT/OzDES survey.

\subsection{IAU SN designation criteria (presented by Seppo Mattila)}

As of January 1, 2016, the Transient Name Server (hereafter TNS) serves as the official IAU instrument for reporting astronomical transient candidates. 
Currently a transient can only obtain an IAU supernova designation (e.g., SN~2018xx)  once it is spectroscopically typed.
This criteria is a continuation of the  scheme previously adopted by the Central Bureau for Astronomical Telegrams (CBETs) and is rooted in a time when there was essentially 
no ability to find highly-extinguished supernovae  in searches limited to optical wavelengths.
 
 However, today  technology has transformed the way in which some of us search for supernovae, and as a result, the current naming criteria is not always adequate.  
As pointed out by S. Mattila, over the past several years making use of laser guide star adaptive optics in the near-IR, the SUNBIRD collaboration  has discovered a number of core-collapse supernovae in luminous infrared galaxies (e.g., \cite{kool18}). However, due to  high extinction it is often not possible to obtain clear optical detections, let alone a visual-wavelength spectrum to secure an IAU supernova designation.  
Furthermore, near-IR spectroscopy is often equally challenging for faint supernovae against the bright and crowded nuclear regions of luminous infrared galaxies.

Near-IR and/or radio follow-up can confirm the supernova nature for such events with no doubt even in the absence of spectroscopic confirmation.  
As an example of such an object, S. Mattila pointed out the case of SN~2008iz (among others), which occurred in the very nearby starburst galaxy M82 and was discovered at radio wavelengths (\cite{brunthaler09}). Although this object had no detected optical emission, adaptive optics $K$-band imaging obtained with  Gemini-N provided a  detection at late times (\cite{mattila13}).  
Very Large Array  provided radio light curves consistent with a core-collapse supernova and Very Long Baseline Interferometry  radio imaging revealed an expanding shell confirming with no doubt that SN~2008iz was a core-collapse supernova  
(see \cite{kimani16}). 

After a heated debate on this subject there appeared to be a general consensus  among the workshop participants that the IAU supernova working group should consider to modify the current IAU criteria for official supernova name designation.  In doing so, the IAU  can ensure that all supernovae discovered within our own galactic neighborhood are able to obtain the IAU designation that they rightfully deserve.

\end{document}